\documentclass[aps,pra,reprint,twocolumn,showpacs,floatfix]{revtex4-2}
\usepackage{amsmath,amssymb}
\usepackage{graphicx}
\usepackage{braket}
\usepackage{booktabs}

\newcommand{\tr}{\operatorname{Tr}}
\newcommand{\nth}{n_{\mathrm{th}}}
\newcommand{\eps}{\varepsilon}

\begin{document}

\title{Role of the drive in mediating correlations between two qubits through a shared dissipative cavity}

\author{Nozhat Ghaseminezhad}
\author{Vahid Ameri}
\email[]{vahameri@gmail.com}
\affiliation{Department of Physics, Faculty of Science, University of Hormozgan, Bandar Abbas, Iran}
\author{Alidad Askari}
\affiliation{Department of Physics, Faculty of Science, University of Hormozgan, Bandar Abbas, Iran}

\date{\today}

\begin{abstract}
Two qubits coupled only through a shared, damped, driven cavity can become correlated even without a direct interaction. Using a numerically exact master equation, we show that the cavity drive sets both the strength and the character of the resulting correlations. Parametric, coherent, and resonantly modulated drives generate quantum discord that grows with cavity temperature, while the logarithmic negativity remains numerically zero: discord without entanglement. A time-modulated parametric drive is the exception, producing genuine two-qubit entanglement with \(E_\mathcal{N} \simeq 0.15\) and concurrence \(\simeq 0.16\) at \((\eps, \gamma) = (0.3, 0.2)\), rising to \(E_\mathcal{N} \simeq 0.32\) in the weak-coupling, moderate-damping region. This entanglement vanishes near \(n_{\mathrm{th}} \simeq 0.2\) as the cavity heats, while the discord keeps increasing, so a single drive delivers a temperature-tuned transition from entanglement to discord. The parametric drive also best preserves coherence. An adiabatic-elimination model attributes the discord without entanglement to the temperature-enhanced effective coupling and collective dephasing generated by the cavity.
\end{abstract}

\maketitle

\section{Introduction}

Entanglement is only one kind of quantum correlation. Quantum discord \cite{Ollivier2001,Henderson2001,Modi2012,Datta2008}
captures non-classical correlations beyond entanglement, and it can persist, or even
appear from nothing, in regimes where entanglement is strictly absent
\cite{Werlang2010,Ferraro2010,Zurek1982}. Two uncoupled subsystems that share a common environment make this
especially transparent: the shared reservoir acts as a correlation mediator, generating
quantum correlations with no direct coupling between the subsystems \cite{Yuan2010,Braun2002,Plenio2002,Benatti2003,WangCav2011,WangChen2013}.
\par
We study two qubits with no direct coupling, interacting only through a common, damped,
driven optical cavity. That a shared environment can create and amplify quantum correlations
between uncoupled subsystems is well established \cite{Braun2002,Yuan2010}; our aim is not to
revisit it, but to ask how the specific cavity drive controls not only how much correlation
appears but what kind. Parametric, coherent, and resonantly modulated drives all give
temperature-growing discord without entanglement, a known common-environment effect. A fourth
drive, a parametric time-modulated pump resonant at $\omega=\Delta$, behaves differently: it is
the only scheme that produces genuine two-qubit entanglement at zero temperature, and heating
then converts that entanglement into discord. The drive also sets how temperature reshapes the
balance between discord and coherence. Cavity drives acting on qubits have been studied before
\cite{Cecoi2020,Xiao2010}, but not, to our knowledge, as a controlled comparison of which drive
yields entanglement rather than discord alone.
\par
We set up the model in Sec.~\ref{sec:model}, examine the static drives in
Secs.~\ref{sec:temp} and \ref{sec:sweep}, treat the pulsed parametric pump in
Sec.~\ref{sec:modpar}, and develop the effective two-qubit model in Sec.~\ref{sec:analytic}.

\section{Model and methods}\label{sec:model}

\begin{figure}[tp]
\includegraphics[width=0.8\columnwidth]{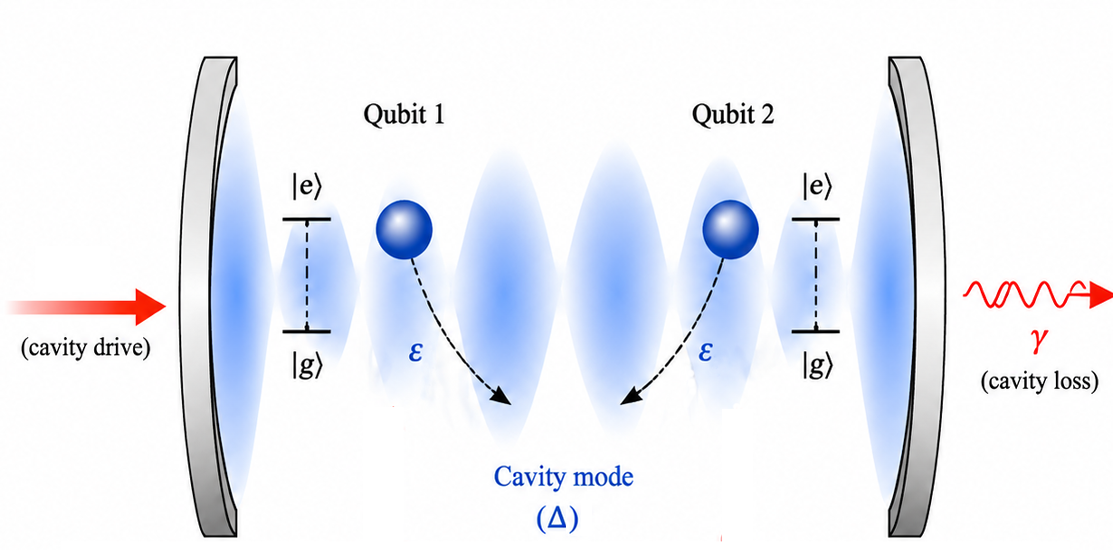}
\caption{\label{fig:tup} Two distinct qubits are coupled to a shared cavity.}
\end{figure}
We consider two two-level subsystems (qubits, indices $1$ and $2$) interacting with a single
cavity mode $3$. The qubits are described by Pauli operators
$\sigma_x^{(i)}, \sigma_y^{(i)}, \sigma_z^{(i)}$ with transition frequency
$\omega_1=\omega_2=\omega$, and the cavity by annihilation operator $a_3$ with frequency
$\omega_3=\omega$. We work in the interaction picture (rotating frame), so that the free part of the Hamiltonian reduces to the detunings
\begin{equation}
  H_0 = \sum_{i=1}^{2}\Delta_i\,\sigma_z^{(i)} + \Delta_3\, a_3^\dagger a_3,
  \label{eq:H0}
\end{equation}
where $\Delta_i$ and $\Delta_3$ are the qubit and cavity detunings from the rotating-frame
reference. We choose a finite, common detuning $\Delta_i=\Delta_3=\Delta=1$
so that the modes are driven off resonance. The parametric and coherent drives are static
operators in this frame, while the modulated drive is tuned to resonance. The qubits interact with the cavity through a linear quadrature coupling,
\begin{equation}
  H_{\mathrm{int}} = \eps\,(\sigma_x^{(1)}+\sigma_x^{(2)})(a_3+a_3^\dagger),
  \label{eq:Hi}
\end{equation}
where $\sigma_x^{(i)}=\sigma_+^{(i)}+\sigma_-^{(i)}$ is the qubit quadrature, with
$\sigma_\pm^{(i)}=(\sigma_x^{(i)}\pm i\sigma_y^{(i)})/2$, and $\eps$ is the common
qubit-cavity coupling. There is no direct qubit-qubit coupling: the two qubits interact only
through the shared cavity mode.
\par
The rotating-wave approximation has not been used in Eq.~\eqref{eq:Hi}. The counter-rotating terms $\sigma_+^{(i)} a_3^\dagger + \sigma_-^{(i)} a_3$ are
retained because the normalized coupling $\eps/\omega=0.3$ lies in the strong-coupling
regime, where these terms are non-negligible and contribute to the correlations.
We compare four standard driving schemes for optical cavities in quantum optics
\cite{Scully1997,Walls1994,Yurke1986}. In the rotating frame, the static drives appear as
static operators,
\begin{itemize}
\item Parametric drive:
  $H_{\mathrm{drive}} = E\,(a_3^{\dagger 2} + a_3^2)$ with $E=0.3\,\omega$;
\item Coherent drive:
  $H_{\mathrm{drive}} = E_1\,(a_3^\dagger + a_3)$ with $E_1=0.4\,\omega$;
\item Modulated drive:
  $H_{\mathrm{drive}} = A\sin^2(w t)\,\omega\,(a_3+a_3^\dagger)/\sqrt{2}$,
  with $A=1$ (dimensionless) and $w=0.5\,\omega$;
\item Time-modulated parametric drive:
  $H_{\mathrm{drive}} = A_{\mathrm{mp}}\sin^2(w_{\mathrm{mp}} t)\,(a_3^{\dagger 2}+a_3^2)$,
  with $A_{\mathrm{mp}}=0.10\,\omega$ and $w_{\mathrm{mp}}=\Delta=\omega$.
\end{itemize}
The first three schemes are the typical drives studied in Secs.~\ref{sec:temp}--\ref{sec:sweep};
the fourth, a pulsed parametric pump, is the central subject of
Sec.~\ref{sec:modpar}. Because $\sin^2(wt)=\tfrac12[1-\cos(2wt)]$, the modulated drive pumps
the cavity at frequency $2w$. We set $2w=\Delta=1$, i.e.\ $w=\Delta/2=0.5$, so that the
modulation is resonant. It resonantly displaces the detuned cavity mode and maximizes
the shared cavity field available as a correlation mediator for the qubits. The fourth
drive, by contrast, is a two-photon process. It is resonant when
$2w_{\mathrm{mp}}=2\Delta$, i.e.\ $w_{\mathrm{mp}}=\Delta=1$ (twice the modulated-drive
frequency), and we work at this resonance.
We choose the parametric amplitudes below the threshold $E=\omega/2$ where the Fock
truncation of the cavity becomes ineffective, keep
$A_{\mathrm{mp}}=0.10$ below its parametric-instability threshold, and set the coherent and
modulated drives so that the three static schemes operate at comparable cavity occupation
$\langle a_3^\dagger a_3\rangle \sim 1$ (the fourth drive, a two-photon pump, naturally
operates at a lower occupation $\langle a_3^\dagger a_3\rangle \sim 0.2$).
\par
The cavity is damped at rate $\gamma$ to a thermal bath of mean occupation $\nth$,
described by the Lindblad dissipators,
\begin{equation}
  \mathcal{L}[\rho] = \gamma(\nth+1)\,D[a_3]\rho + \gamma\nth\, D[a_3^\dagger]\rho,
\end{equation}
where $D[O]\rho = O\rho O^\dagger - \tfrac12\{O^\dagger O,\rho\}$, within the standard
Lindblad formulation of Markovian open quantum dynamics
\cite{Lindblad1976,Gorini1976,Breuer2002}. The qubits have no intrinsic dissipation. Solving the master equation
$\dot\rho = -i[H,\rho] + \mathcal{L}[\rho]$ numerically with QuTiP \cite{Qutip} and
using a Fock-truncated cavity of dimension $N_3=20$, we derive the two-qubit reduced state
$\rho_{12} = \tr_3 \rho$.

\par
Without qubit dissipation, the Liouvillian has a
non-unique steady-state manifold, so the reduced two-qubit state at long times depends on the
initial preparation. We always specify the (physical) long-time
state reached from the separable initial state $\ket{000}$ (all modes in vacuum). This state is
fully converged in time (the two-qubit discord is unchanged
from $t\sim 500$ to $t\sim 2000$). The absolute values are initialization-dependent, while the
comparisons between driving schemes are made on equal footing from the same protocol.
As we verify in the discussion (Sec.~\ref{sec:concl}), the central conclusions, in
particular the pulsed-drive entanglement, survive the initial preparation and
the introduction of a small intrinsic qubit decay, which also removes the
non-uniqueness of the steady state.
\par
From $\rho_{12}$ we compute,
\begin{itemize}
\item the quantum discord $\mathcal{D}$, defined as the difference between the
  mutual information and the classical correlations,
  $\mathcal{D} = \mathcal{I} - \mathcal{C}$, with
  $\mathcal{I} = S(\rho_1)+S(\rho_2)-S(\rho_{12})$ and the classical correlations maximized over
  projective measurements \cite{Ollivier2001,Henderson2001}.
\item the coherence, quantified by the $\ell_1$-norm coherence
  $C_{\ell_1} = \sum_{i\neq j}|\rho_{ij}|$ \cite{Baumgratz2014}.
\end{itemize}

\section{Temperature enhancement of discord and scheme-dependent coherence}\label{sec:temp}
\begin{figure}[ht]
\includegraphics[width=\columnwidth]{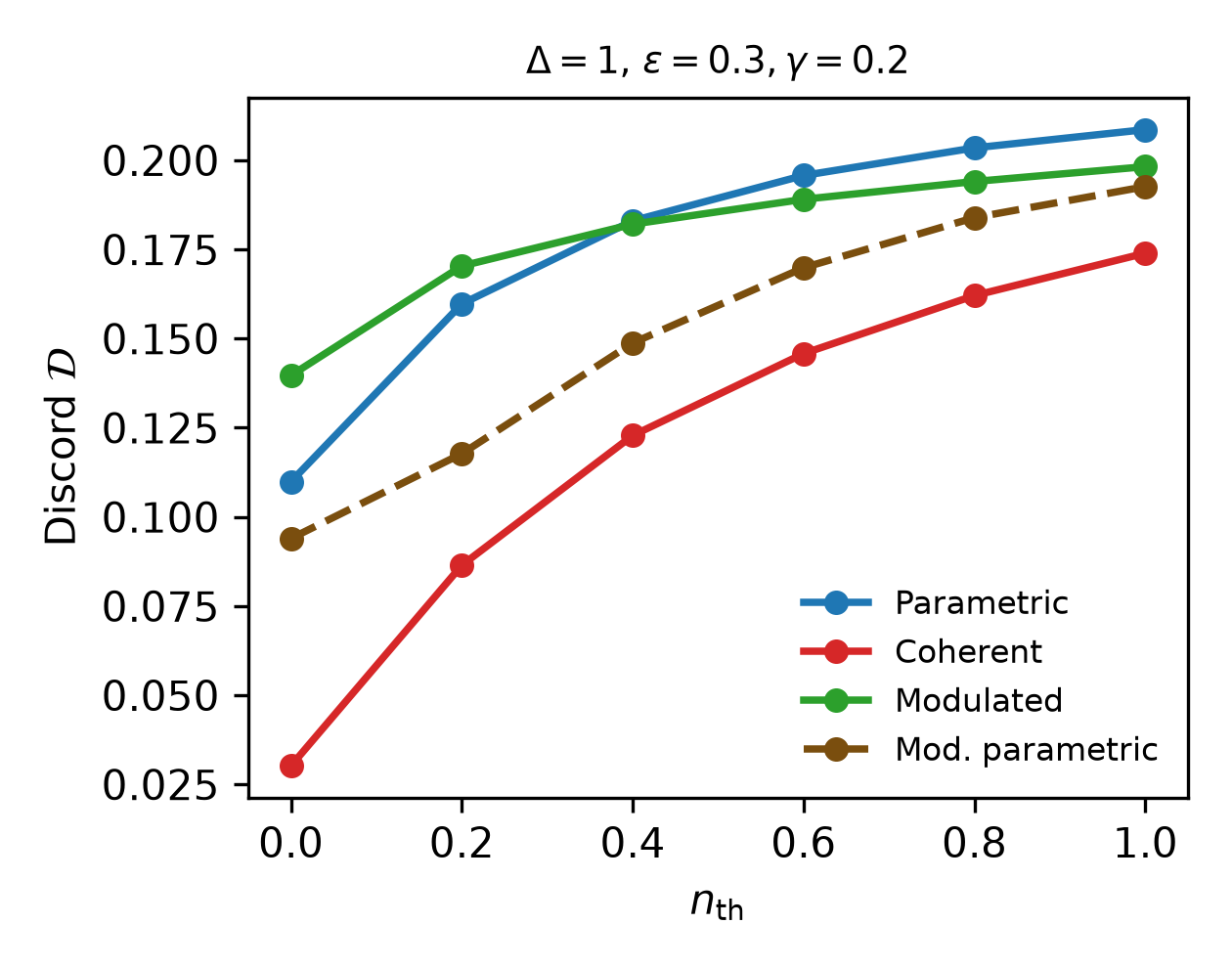}
\caption{\label{fig:temp} Steady-state quantum discord $\mathcal{D}$ as a function of the cavity
thermal occupation $\nth$ for the parametric (blue), coherent (red), modulated (green), and
time-modulated parametric (dashed brown) drives, at detuning $\Delta=1$, coupling $\eps=0.3$,
damping $\gamma=0.2$ ($N_3=20$).}
\end{figure}

Figure~\ref{fig:temp} shows the steady-state discord against cavity temperature
$\nth$ for all four drives. The discord grows with temperature in every case, saturating at
high $\nth$. For the three static drives (parametric, coherent, and modulated) the growth is
discord without entanglement: the logarithmic negativity
$E_\mathcal{N}$ \cite{Vidal2002}, a full entanglement monotone although not convex
\cite{Plenio2005}, is numerically zero. The coherent drive shows only a tiny residual
($\sim 10^{-3}$) at zero temperature, which vanishes on heating, so at finite temperature
$E_\mathcal{N}$ is zero within numerical precision. Only the pulsed drive
produces entanglement at zero temperature, and it is destroyed on heating
(Sec.~\ref{sec:modpar}). We treat $E_\mathcal{N}$ for the static drives as a witness
that the cavity-generated correlations are essentially non-entangled, not as a resource, and we
attach no meaning to the residual round-off.

For two qubits correlated only through a common
damped cavity, thermal noise enhances the discord. This
is consistent with the known ability of a common environment to generate and amplify quantum
correlations without entangling the subsystems \cite{Yuan2010,Braun2002,Plenio2002,Benatti2003,
Paz2002,Kim2010,Wang2011,Obada2009,WangCav2011,Restrepo2016,Berrada2011,XuLi2005}.
\par
\begin{figure}[ht]
\includegraphics[width=\columnwidth]{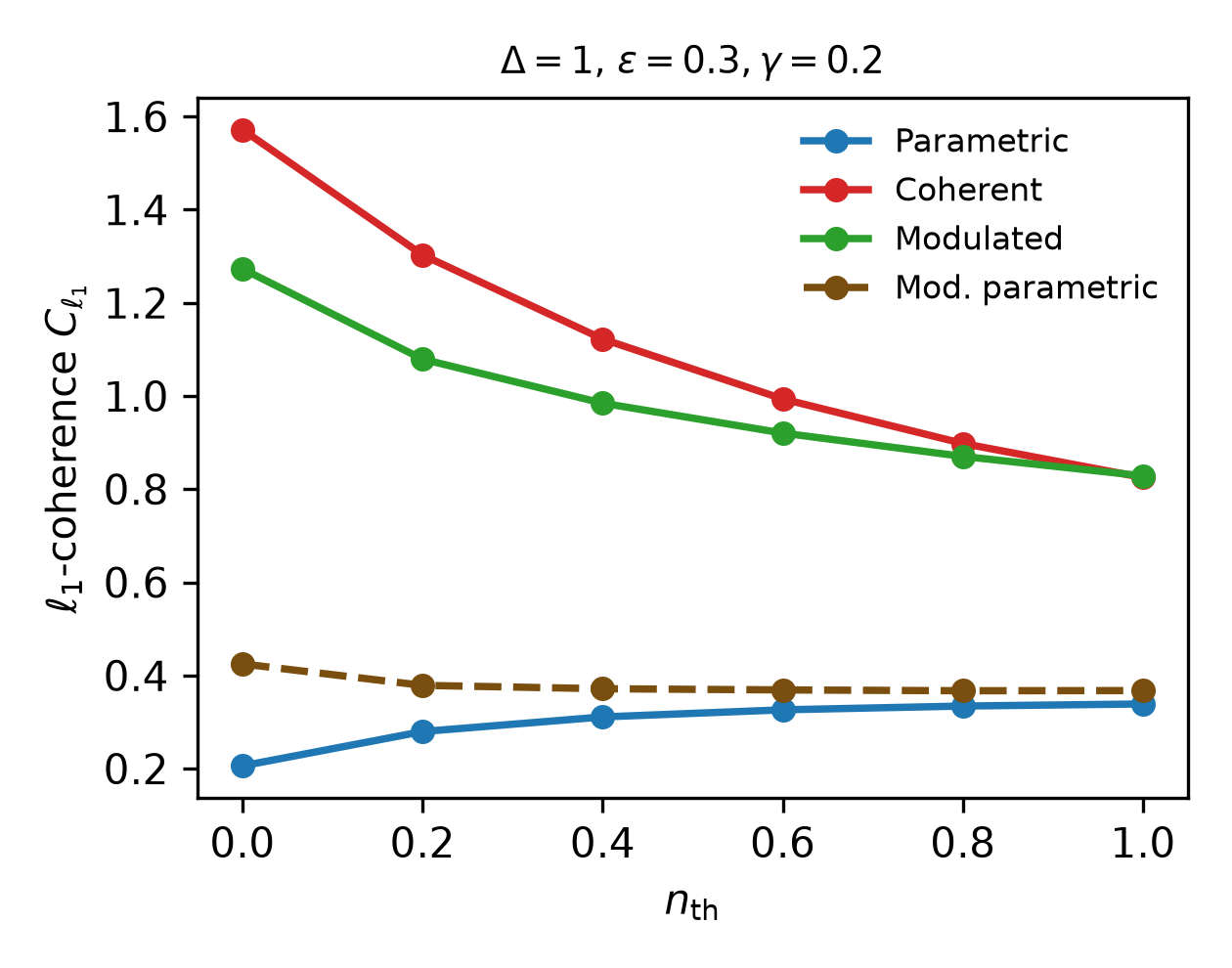}
\caption{\label{fig:coherence} $\ell_1$-norm coherence of the two-qubit state as a function of
$\nth$ for the parametric (blue), coherent (red), modulated (green), and time-modulated
parametric (dashed brown) drives at $\Delta=1,\ \eps=0.3,\ \gamma=0.2$ ($N_3=20$).}
\end{figure}

The coherence behaves differently (Fig.~\ref{fig:coherence}):
\begin{itemize}
\item Parametric: $C_{\ell_1}$ rises with temperature ($0.21\to0.34$)
  while the discord simultaneously grows ($0.11\to0.21$). This drive converts thermal noise into
  both discord and coherence, and holds up best under heating in this respect.
\item Coherent: $C_{\ell_1}$ collapses with temperature ($1.57\to0.83$) as the
  discord grows ($0.03\to0.17$).
\item Modulated: $C_{\ell_1}$ falls ($1.27\to0.83$), and the discord grows
  ($0.14\to0.20$).
\item Time-modulated parametric: $C_{\ell_1}$ stays roughly flat ($0.42\to0.37$) while the
  discord grows ($0.09\to0.19$), and at zero temperature it additionally carries genuine
  entanglement (Sec.~\ref{sec:modpar}).
\end{itemize}
Temperature thus plays a dual role: it builds discord while
destroying coherence in the coherent and modulated cases. The
parametric drive is unusual in protecting and enhancing coherence while the discord grows.
This scheme-controlled decoupling of coherence from discord is one of our main results.

\section{Parameter study: coupling and damping}\label{sec:sweep}
\begin{figure}[!t]
\includegraphics[width=\columnwidth,height=0.55\textheight,keepaspectratio]{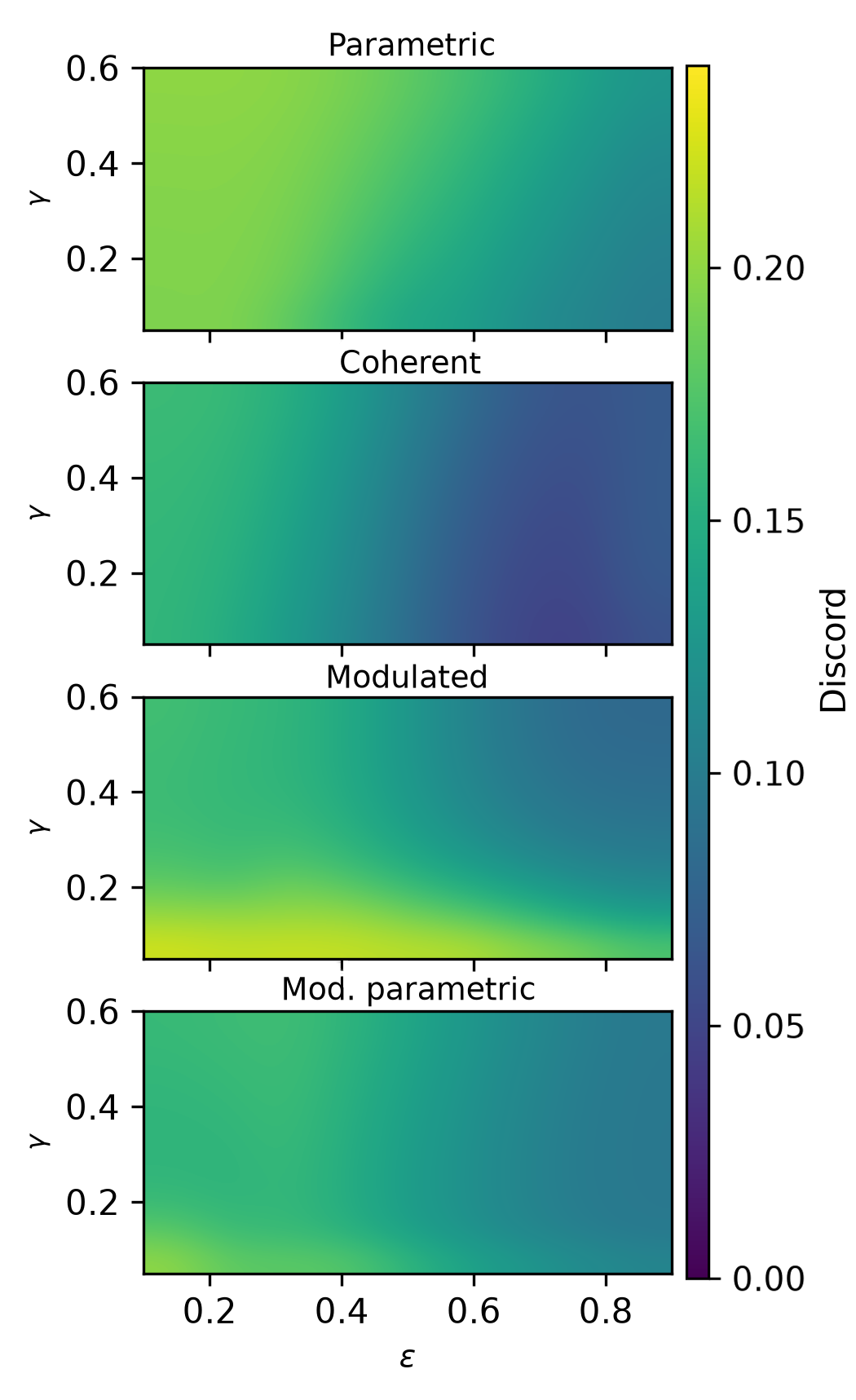}
\caption{\label{fig:phasediagram} Steady-state discord in the $(\eps,\gamma)$ plane at
$\nth=0.5$ (9$\times$9 grid) at $\Delta=1$, for the parametric, coherent, modulated, and
time-modulated parametric (mod.\ param.) drives.}
\end{figure}

To map the operating points, we swept the qubit-cavity coupling $\eps$ and the cavity damping
$\gamma$ at fixed $\nth=0.5$. Figure~\ref{fig:phasediagram} shows the steady-state discord
as a color map over the $(\eps,\gamma)$ plane for all four schemes.

\par
For the parametric and coherent schemes, the discord is
maximized at weak coupling and decreases as $\eps$ grows, while the coherence (not
shown) increases with $\eps$. The modulated drive, by contrast, is weakly non-monotonic
in $\eps$ and remains relatively large across a broad range of couplings. The coupling
provides a knob that, for these two schemes, trades discord against
coherence, while for the modulated drive it defines a broad optimal operating region.

\par
Here the schemes respond
oppositely: the modulated discord decreases with
$\gamma$, whereas the parametric and coherent discord increase with $\gamma$. The modulated
drive is therefore favored at low loss, and its
global maximum ($0.222$ at
$\eps\simeq0.10$, $\gamma=0.05$; Fig.~\ref{fig:phasediagram}) exceeds the maximum discord
reached by the parametric ($0.199$) or coherent ($0.164$) drives anywhere in the studied
plane, occurring at weak coupling and low loss.

Both parameter studies show that the balance between discord and coherence is controllable in
the driven common cavity: the coupling sets the mix of the two, and the damping selects the
best drive. The modulated drive reaches the highest discord of any scheme, at weak coupling and
low loss, while the parametric and coherent schemes do better at weak coupling and high loss.

\section{Fourth scheme: time-modulated parametric pump}\label{sec:modpar}

The fourth scheme makes the parametric pump time-periodic:
\begin{equation}
  H_{\mathrm{drive}} = A\sin^2(w t)\,(a_3^{\dagger 2}+a_3^2),
\end{equation}
in the same rotating frame as Sec.~\ref{sec:model}. Because $\sin^2(wt)=\tfrac12[1-\cos(2wt)]$
contains a component at $2w$ and the parametric pump is a two-photon process, the drive
is resonant when $2w=2\Delta$, i.e.\ $w=\Delta$. This is twice the frequency of the
modulated drive (single-photon, $w=\Delta/2$). A scan of the modulation frequency
confirms the cavity occupation peaks at $w\simeq\Delta$. We work at $w=\Delta=1$.

As with the static parametric pump, this scheme has a threshold for parametric instability, beyond which the Fock truncation becomes ineffective. We have confirmed that the results are converged for
$A\lesssim0.10$ ($\langle n\rangle$ is unchanged between $N_3=20$ and $30$), and use
$A=0.10$.

At low temperature this scheme behaves differently. At $\nth=0$ and the reference operating
point ($\eps=0.3,\gamma=0.2$), the two-qubit reduced state is entangled, with a logarithmic
negativity of $E_\mathcal{N}\simeq 0.15$ and a concurrence of
$\mathcal{C}\simeq 0.16$ (period-averaged values), fluctuating between about $0.09$ and $0.18$
over a drive period; the values are converged in both $N_3$ and time. The three static schemes
do not entangle the qubits: their log-negativity remains numerically zero, apart from the
$\sim 10^{-3}$ coherent-drive residual at zero temperature that vanishes on heating
(Sec.~\ref{sec:temp}).

The time-modulated two-photon pump is the only one of the four schemes that escapes the
discord-without-entanglement regime \cite{Zheng2014,Nourmandipour2015}. Recent work has studied entanglement between two qubits in
a driven common cavity \cite{Dey2025,Dey2026}, focusing on directly driving the qubits and on
the size of the steady-state entanglement and the coupling asymmetry. We instead vary the
cavity drive itself across four schemes and track which correlations each drive produces and how
temperature changes them, rather than studying gate operations or coupling-ratio thresholds
\cite{Srinivasa2024}. The novel elements are the drive-by-drive comparison and the single-drive,
temperature-tuned transition from entanglement to discord.

The entanglement is not a narrow numerical feature. Scanning the $(\eps,\gamma)$ plane at
$\nth=0$ shows it over a broad region, strongest at weak coupling and moderate
damping, reaching $E_\mathcal{N}\simeq0.32$ at $\eps=0.1,\gamma=0.1$. Entanglement that survives
through the whole drive period occurs for $\eps\lesssim0.3$ and $\gamma$ between about $0.1$ and
$0.5$. It is suppressed at very low damping ($\gamma\simeq0.05$), where the pulsed pump has too
little time to act each period, and at stronger coupling ($\eps\gtrsim0.4$), where the qubit
back-action degrades the two-photon pump. We quote the period-averaged value at the reference
point ($\eps=0.3,\gamma=0.2$) and, separately, the plane maximum
($E_\mathcal{N}\simeq0.32$ at $\eps=0.1,\gamma=0.1$) cited in the abstract.

\begin{figure}[!t]
\includegraphics[width=\columnwidth,keepaspectratio]{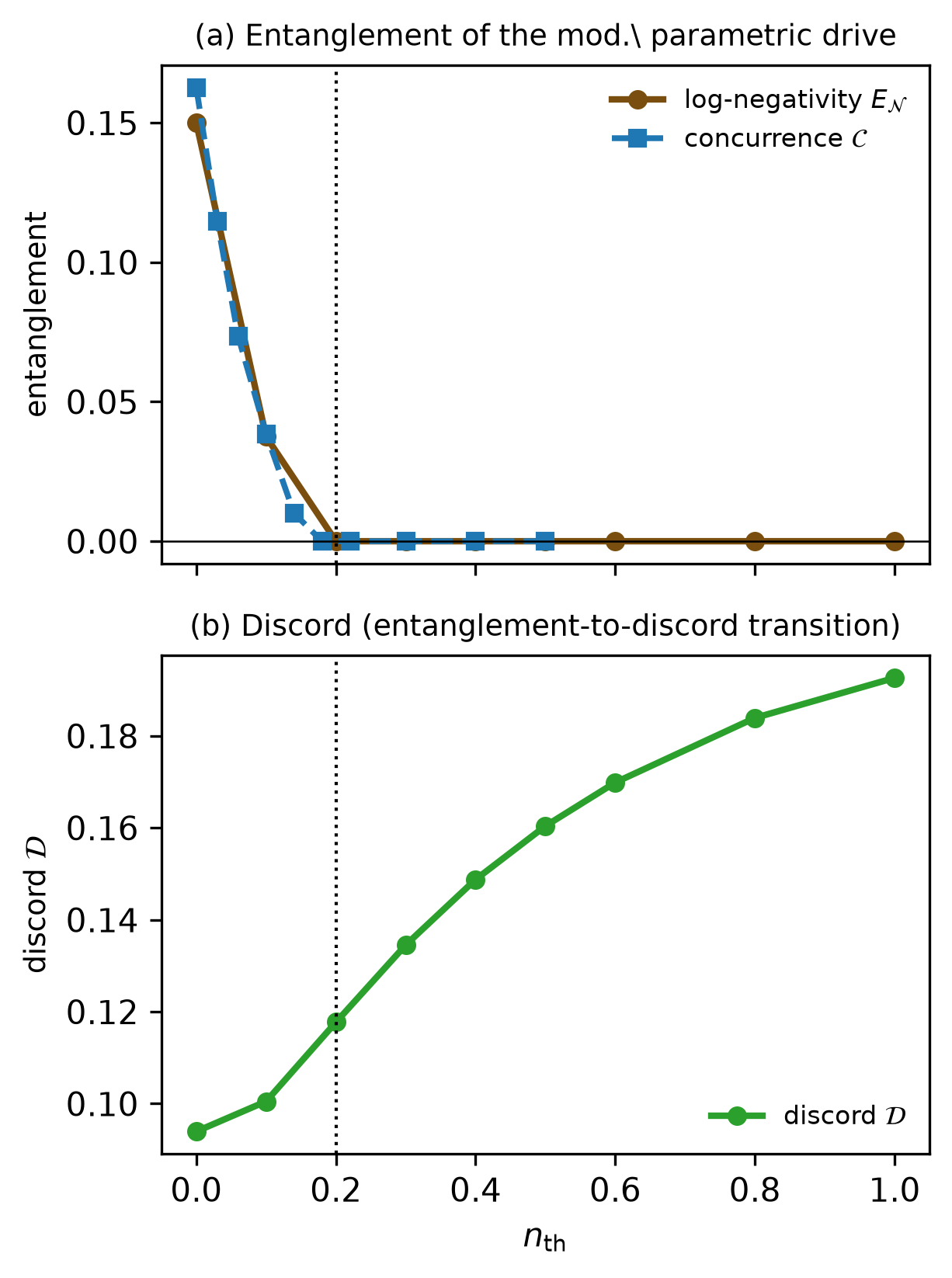}
\caption{\label{fig:modpar_ent} Entanglement-to-discord transition of the time-modulated
parametric drive ($A=0.10,\ w=\Delta=1$, $\eps=0.3$, $\gamma=0.2$, $N_3=20$).
(a) Log-negativity $E_\mathcal{N}$ (solid) and concurrence $\mathcal C$ (dashed) versus the
cavity temperature $\nth$: the two-qubit entanglement is destroyed by $\nth\simeq0.2$.
(b) The quantum discord, which keeps growing as the entanglement vanishes, so that a single drive
tunes the correlations from entangled at low temperature to discordant (still non-classical)
at high temperature.}
\end{figure}

Heating destroys the entanglement while the discord grows, giving a clean entanglement-to-discord crossover (Fig.~\ref{fig:modpar_ent}). The entanglement measure \(E_\mathcal{N}\)
decreases steadily and vanishes around \(\nth \simeq 0.2\), while the discord rises to about
\(\mathcal{D} \simeq 0.19\) at \(\nth = 1\), comparable to the modulated drive but at a much
smaller amplitude (\(A = 0.10\)). The \(\ell_1\) coherence stays moderate and nearly constant,
from about \(0.37\) to \(0.42\), between the rising parametric case and the collapsing coherent
case. The fourth scheme combines the two regimes of Sec.~\ref{sec:temp}: entanglement
at low temperature and discord without entanglement at high temperature.
\par
The entanglement comes from resonant two-photon dynamics. The modulated pump, \(\sin^2(wt)(a_3^{\dagger 2} + a_3^2)\), switches on and off within the drive period, acting each cycle as a pulsed two-photon pump at the cavity frequency. Degenerate two-photon (parametric) processes generate nonclassical two-photon correlations within a single mode \cite{Yuen1976}, and cavity two-photon processes are known to drive entanglement, including its finite-time death \cite{Cardoso2009,Khalil2010}. The cavity transmits this nonclassicality to the qubits as a two-photon-excitation correlation.
\par
The static parametric drive, by contrast, acts continuously and stabilizes the detuned cavity in a steady state whose pair correlations are weakened by the balance of drive and loss, so it cannot entangle the qubits (Sec.~\ref{sec:temp}).
\par
The pulsed nature of the pump is what matters: two-photon correlations are injected intermittently and used before the loss channel can decorrelate them. The entanglement is strongest at moderate damping, \(\gamma \simeq 0.1\) to \(0.5\). The pump needs enough loss to reach a steady state, but not so much that the pulsed two-photon correlation is destroyed within a single period.

\section{Effective two-qubit model}\label{sec:analytic}

The numerics can be understood through a minimal analytic model obtained by adiabatically eliminating the fast, strongly damped cavity. Consider the coupling
$H_{\mathrm{int}} = \eps\,(\sigma_x^{(1)}+\sigma_x^{(2)})(b+b^\dagger)$ to a cavity mode $b$ damped at rate
$\gamma$. The cavity position quadrature $X = b+b^\dagger$ has steady-state mean
$\langle X\rangle = \alpha+\alpha^*$ (drive displacement) and variance
$\mathrm{Var}(X) = 1 + 2 n_{\mathrm{eff}}$, where $n_{\mathrm{eff}}$ contains both the thermal
and the drive-induced occupancy. Eliminating $b$ gives an effective model that includes, in addition to the Ising exchange term, a linear drive term arising from the nonzero cavity displacement,
\begin{equation}
\begin{aligned}
  H_{\mathrm{eff}} &= h\,(\sigma_x^{(1)}+\sigma_x^{(2)}) + J\,\sigma_x^{(1)}\sigma_x^{(2)},\\[2pt]
  h &\sim \eps\langle X\rangle,\qquad
  J \sim \frac{2\eps^2}{\Delta},
  \label{eq:eff}
\end{aligned}
\end{equation}
together with a collective dephasing channel. Because the coupling is via the $x$
quadrature, the collective dephasing acts in the same basis,
$\Gamma\,(\sigma_x^{(1)}+\sigma_x^{(2)})$ type, with $\Gamma \sim \eps^2\,\mathrm{Var}(X)$. The collective
dephasing strength $\Gamma$ grows with the cavity field (photon number), and it correlates the two qubits through the shared reservoir without entangling them.
\par
Because $\nth$ raises the cavity variance $\mathrm{Var}(X)=1+2n_{\mathrm{eff}}$, temperature increases $\Gamma$, enhancing the shared correlation and hence the discord. For the
coherent drive the coherence is set by $|\alpha|$, and the growing thermal occupancy washes this out, which explains the coherence collapse in the coherent and modulated cases.
\par
To test this mechanism, we solve the effective model with the coupling $J$,
\begin{equation}
  \dot\rho = -i[H_{\mathrm{eff}},\rho] + \Gamma\,D[\sigma_x^{(1)}+\sigma_x^{(2)}]\rho,
  \label{eq:effmaster}
\end{equation}
where a weak local relaxation term is included only to ensure a definite steady state for this model. The full numerical analysis does not require such a term, since the common cavity provides the only source of dissipation.
\par
The discord increases with \(J\), peaks at moderate \(J\), and then declines, while the
logarithmic negativity stays at zero throughout. This reproduces the discord-without-entanglement
behavior of the three static drives. The minimal model is a qualitative argument, not a
quantitative fit: it identifies the common-cavity effective coupling and collective dephasing as
the source of the discord without reproducing the full numerical values.
\par
The static-drive model contains only single-excitation (Ising and linear-drive) terms, which couple the qubits through a classical channel and cannot entangle them. Capturing the fourth scheme requires the coherent two-photon (pair) coupling mediated by the pulsed pump. Diagonalizing the pulsed-drive two-qubit reduced state confirms this directly: its entanglement lies entirely between the $\ket{00}$ and $\ket{11}$ states (the off-diagonal element $\rho_{03}$ is nonzero), a two-mode-squeezed, down-conversion structure, the direct two-qubit analog of a degenerate two-photon (parametric) process \cite{Yuen1976}. The pulsed pump drives an oscillating cavity two-photon amplitude $\langle a_3^2(t)\rangle$ at $2\Delta$, resonant with the two-qubit pair transition, whereas the static parametric drive gives a constant $\langle a_3^2\rangle$ whose pair correlations are suppressed by the balance of drive and loss. The minimal reduced Hamiltonian that closes the mechanism is
\begin{equation}
  H_{\mathrm{eff}}' = h\,(\sigma_x^{(1)}+\sigma_x^{(2)}) + J\,\sigma_x^{(1)}\sigma_x^{(2)}
  + g_2\,\bigl(\sigma_+^{(1)}\sigma_+^{(2)} + \sigma_-^{(1)}\sigma_-^{(2)}\bigr),
  \label{eq:eff2}
\end{equation}
where the term proportional to $g_2$ couples $\ket{00}\leftrightarrow\ket{11}$ and is the entangling ingredient, together with the same collective dephasing $\Gamma\,D[\sigma_x^{(1)}+\sigma_x^{(2)}]$. For the static drives $g_2=0$ and Eq.~\eqref{eq:eff} is recovered.
\par
We test Eq.~\eqref{eq:eff2} in two ways. First, for a coherent (lossless) two-photon pulse, the pair term alone drives $\ket{00}$ to the maximally entangled Bell state $(\ket{00}+\ket{11})/\sqrt2$ after a $\pi/2$ pulse, with logarithmic negativity $\ln 2\simeq0.693$ and concurrence $1$. Second, the cavity-induced collective dephasing $D[\sigma_x^{(1)}+\sigma_x^{(2)}]$ destroys this entanglement while leaving a finite discord, as in the full model: the log-negativity falls from $0.557$ to $0.10$ as $\Gamma$ rises from $0.05$ to $1.0$, while the discord survives. The same two-photon reduced model reproduces both the low-dephasing (low-temperature) entanglement and its conversion to discord at high dephasing (high temperature), i.e.\ the entanglement-to-discord transition of Sec.~\ref{sec:modpar}. Again, this is a qualitative, mechanistic argument rather than a quantitative fit to the full master equation.

\section{Discussion and conclusions}\label{sec:concl}
The way a common damped cavity is driven determines the type of correlation that develops
between two otherwise uncoupled qubits. Of the schemes we studied, only the pulsed
parametric pump (Sec.~\ref{sec:modpar}) yields genuine two-qubit entanglement:
\(E_\mathcal{N}\simeq0.15\) and concurrence \(\simeq0.16\) at zero temperature at the reference
point \((\eps,\gamma)=(0.3,0.2)\), reaching \(E_\mathcal{N}\simeq0.32\) in the weak-coupling,
moderate-damping region.
\par
On heating, the entanglement vanishes around \(n_{\mathrm{th}} \simeq 0.2\) while the discord
keeps growing, so a single drive converts entanglement into discord. This differs
from the well-documented ability of a shared environment to generate discord without
entanglement \cite{Yuan2010, Braun2002}.
\par
For the static drives the balance depends on the scheme. Parametric, coherent, and resonantly
modulated drives all give discord that grows with temperature while the correlations stay
unentangled. At the reference point (\(\eps=0.3, \gamma=0.2\)) the parametric drive gives the
most discord of the three and also increases coherence, whereas the coherent and
modulated drives convert noise into discord at the cost of coherence. Coupling and damping
control this balance, which the reduced model attributes to the common-cavity
coupling and a collective dephasing channel.
\par
Taken together, the results give a way to engineer and preserve quantum correlations in a driven
common cavity. They also bear on environment-induced quantum correlations
\cite{Yuan2010,Braun2002}, coherence-discord complementarity
\cite{Baumgratz2014,Streltsov2017,Yu2016,Singh2015}, and the interplay of coherence and
nonclassical correlations in thermal or squeezed reservoirs \cite{Pathania2026,Peng2024}. A
natural experimental realization is the circuit- or cavity-QED platform
\cite{Blais2004,Wallraff2004,Raimond2001}, where qubits couple to a shared resonator with
controllable drive and temperature.
\par
We retain the counter-rotating terms (strong coupling, $\eps/\omega = 0.3$) but express the dissipation in the bare cavity basis. A fully dressed-state treatment would analyze this in the polariton basis \cite{Lednev2024,Costa2026,Settineri2018}, within the broader framework of strong-coupling quantum optics \cite{Kockum2019,DeLiberato2014}. To assess the sensitivity of our results to this choice, we repeated the modulated-parametric calculation using the dissipation expressed in the dressed (polariton) basis. This basis is constructed from the secular decomposition of the bare cavity operator in the eigenbasis of the coupled system $H_0 + H_i$. The primary entanglement is preserved, and even slightly enhanced, by this change (e.g., $E_\mathcal{N}$ increases from 0.15 to 0.21 and concurrence rises from 0.16 to 0.23 at $\eps = 0.3$, $\nth = 0$), and the entanglement-to-discord transition, where entanglement vanishes around $\nth \simeq 0.2$, is unchanged. The bare-basis Lindblad treatment is reliable for the effects we report, which are largest where the two bases closely agree (in the weak-coupling regime).
\par
The modulated drive is time-periodic, so its "steady-state" quantities are averages over a drive
period and depend mildly on the averaging window. The discord is evaluated on a finite
measurement grid whose size slightly shifts the absolute values, although the trends are
converged. The reduced model is likewise a qualitative representation rather than a
quantitative fit.
\par
Finally, we address the initialization dependence flagged in the model section. The absence of qubit dissipation makes the long-time state part of a non-unique manifold. This does not affect the reported physics. First, introducing a small intrinsic qubit decay $\kappa_q$, which gives the Liouvillian a unique steady state and so removes the ambiguity, preserves the entanglement at the reference point. The log-negativity decreases smoothly from $0.150$ to $0.088$ and the concurrence from $0.163$ to $0.093$ as $\kappa_q$ grows from $0$ to $0.03$, i.e.\ the entanglement survives a decay strength comparable to the cavity damping $\gamma=0.2$. Second, the entanglement also survives changes in the initial preparation. At $\kappa_q=0$ it is present for the $\ket{00}$, $\ket{01}$, and thermal-like cavity initial states, with concurrence $0.163$, $0.241$, and $0.163$, respectively (in fact slightly enhanced for $\ket{01}$). The pulsed-drive entanglement is not an artifact of the idealized no-dissipation, $\ket{000}$ preparation.
\par
None of these caveats changes the main conclusion. A single pulsed parametric pump
shifts the two-qubit correlations in a shared cavity from entangled at low temperature to
discordant, still non-classical, at high temperature. Static drives produce only discord without
entanglement, with a scheme-controlled coherence balance.

\begin{acknowledgments}
This work was supported by the Iran National Science Foundation (INSF) through the Persian Gulf Fund (grant number 11328). We thank the University of Hormozgan for support. We also thank the numerical tool QuTiP \cite{Qutip}, which we used in this work.
\end{acknowledgments}

\bibliographystyle{apsrev4-2}
\bibliography{pra}

\end{document}